\documentclass{article}
\usepackage{spconf,amsmath,amssymb,graphicx,hyperref}
\usepackage{bm}
\usepackage{booktabs}
\usepackage{tikz}
\usetikzlibrary{positioning,arrows.meta,calc}
\usepackage{algorithm}
\usepackage{algpseudocode}
\usepackage{multirow}
\usepackage{pdfsync}
\usepackage{array} 
\newcolumntype{M}[1]{>{\centering\arraybackslash}m{#1}}

\newcommand\bw{\ensuremath{{\bm w}}}
\newcommand\bW{\ensuremath{{\bm W}}}

\newcommand\bh{\ensuremath{{\bm h}}}

\newcommand{\setB}{\mathcal{B}}

\usepackage{subcaption}

\newcommand\bH{\ensuremath{{\bm H}}}

\newcommand\bS{\ensuremath{{\bm S}}}

\newcommand{\Nrf}{N_{\mathrm{RF}}}

\definecolor{orange}{RGB}{255,107,0}

\title{Discrete Diffusion for Port Selection in Multiuser Pinching-Antenna Systems}
\name{
Wenxuan Sun$^{\star}$ \quad
Mingjie Shao$^{\dagger}$ \quad
Yanqing Xu$^{\ddagger}$ \quad
Ya-Feng Liu$^{\S}$
}

\address{
\fontsize{11pt}{12pt}\selectfont $^{\star}$School of Information Science and Engineering, Shandong University, Qingdao, China \\
\fontsize{11pt}{12pt}\selectfont $^{\dagger}$State Key Laboratory of Mathematical Sciences, AMSS, Chinese Academy of Sciences, Beijing, China \\
\fontsize{11pt}{12pt}\selectfont $^{\ddagger}$School of Science and Engineering, The Chinese University of Hong Kong, Shenzhen, China \\
\fontsize{11pt}{12pt}\selectfont $^{\S}$School of Mathematical Sciences, Beijing University of Posts and Telecommunications, Beijing, China \\
\small\texttt{$^{\star}$wenxuansun@mail.sdu.edu.cn, $^{\dagger}$mingjieshao@amss.ac.cn}, \small\texttt{$^{\ddagger}$xuyanqing@cuhk.edu.cn, $^{\S}$yafengliu@bupt.edu.cn}
}

\begin{document}
\ninept
\maketitle
\begin{abstract}
Pinching-antenna systems (PASS) reconfigure wireless channels by activating radiating elements at selected locations along dielectric waveguides.
Port selection is a core issue in improving the performance of multiuser PASS systems.
In this paper, we study joint port selection and precoder design for multiuser sum rate maximization, which leads to a large-scale nonconvex mixed-integer problem coupling discrete port selection with continuous precoder design.
We propose a discrete diffusion method that does not require optimal or near-optimal port selection solutions as training labels and recasts the resulting optimization problem as sampling from a target distribution.
Instead, we learn a potential function from local objective differences between neighboring feasible realizations.
The learned potential is incorporated into a Metropolis--Hastings sampling rule to guide a parallel diffusion process while preserving the port selection constraints.
Simulation results show that the proposed method approaches exhaustive search within a $0.30\%$ sum rate gap while achieving over $150\times$ speedup, and outperforms state-of-the-art methods including greedy and beam search.
In particular, it also achieves an $18.8\times$ speedup over greedy search.
\end{abstract}
\section{Introduction}
\label{sec:intro}

Emerging reconfigurable wireless technologies, including reconfigurable intelligent surfaces (RISs), movable antennas (MAs), fluid antenna systems (FASs), and pinching-antenna systems (PASS), are reshaping the physical layer of sixth-generation (6G) networks.
Among these technologies, pinching-antenna systems (PASS) enable flexible activation of radiating points along dielectric waveguides, providing low-loss signal transport, large-scale spatial reconfigurability, and enhanced line-of-sight (LoS) connectivity \cite{liu2025pass,ding2024flexible,liu2025tutorial}. 
Existing studies on pinching-antenna design mainly follow two lines.
The first optimizes the continuous positions of PAs along the waveguides, typically jointly with beamforming and power allocation \cite{wang2025modeling,xu2025rate,zhang2025uplink}.
The second considers discrete port selection, where PAs are activated at a finite set of preconfigured ports \cite{wang2025antenna,11263923,11165763}.
Compared with continuous positioning, port selection imposes lower hardware requirements due to its simple on/off operation and could be easier to implement in practice \cite{liu2025tutorial,wang2025antenna}.

The PASS port selection problem has attracted increasing research interest, including power minimization via branch-and-bound \cite{11263923}, greedy search for over-the-air computation \cite{cheng2026fastpinchingantennaactivationaircomp}, local search for non-orthogonal multiple access (NOMA) transmission \cite{wang2025antenna,11165763}, and graph neural network (GNN)- and attention-based deep learning methods for point-to-point channel capacity maximization \cite{karagiannidis2025deep}.
A core challenge in these studies is to tackle the resulting nonconvex integer nonlinear programming (INLP) problems, which may further couple discrete port selection with continuous precoder design \cite{11263923}.
The number of feasible port selection realizations grows exponentially with the number of RF chains \cite{cheng2026fastpinchingantennaactivationaircomp}, rendering exhaustive search computationally prohibitive.
In particular, using supervised learning with labeled solutions, the GNN-based method in \cite{karagiannidis2025deep} models the antenna-user system as a graph and demonstrates attractive potential for high-performance port selection.

In this work, we consider sum rate maximization in a multiuser downlink PASS system, which necessitates joint port selection and precoder design.
Our key idea is to recast the resulting optimization problem as sampling from a target distribution and develop a discrete diffusion method without requiring optimal or near-optimal port-selection solutions as training labels. 
To this end, we learn a potential function from local objective differences between neighboring feasible selections.
Meanwhile, we construct the diffusion process directly over the feasible set to preserve the combinatorial constraints, thereby avoiding potential performance loss caused by post-processing constraint projections.
The learned potential is incorporated into a Metropolis--Hastings (MH) sampling rule to guide the parallel diffusion process toward high-quality port selections.
Simulation results show that the proposed method approaches exhaustive search within a $0.30\%$ sum rate gap while achieving over $150\times$ speedup, and outperforms state-of-the-art methods including greedy and beam search.
In particular, the proposed diffusion method also achieves an $18.8\times$ speedup over greedy search.

\section{System Model and Problem Formulation}
\label{sec:system}

\begin{figure}[ht]
    \centering
    \includegraphics[width=\columnwidth]{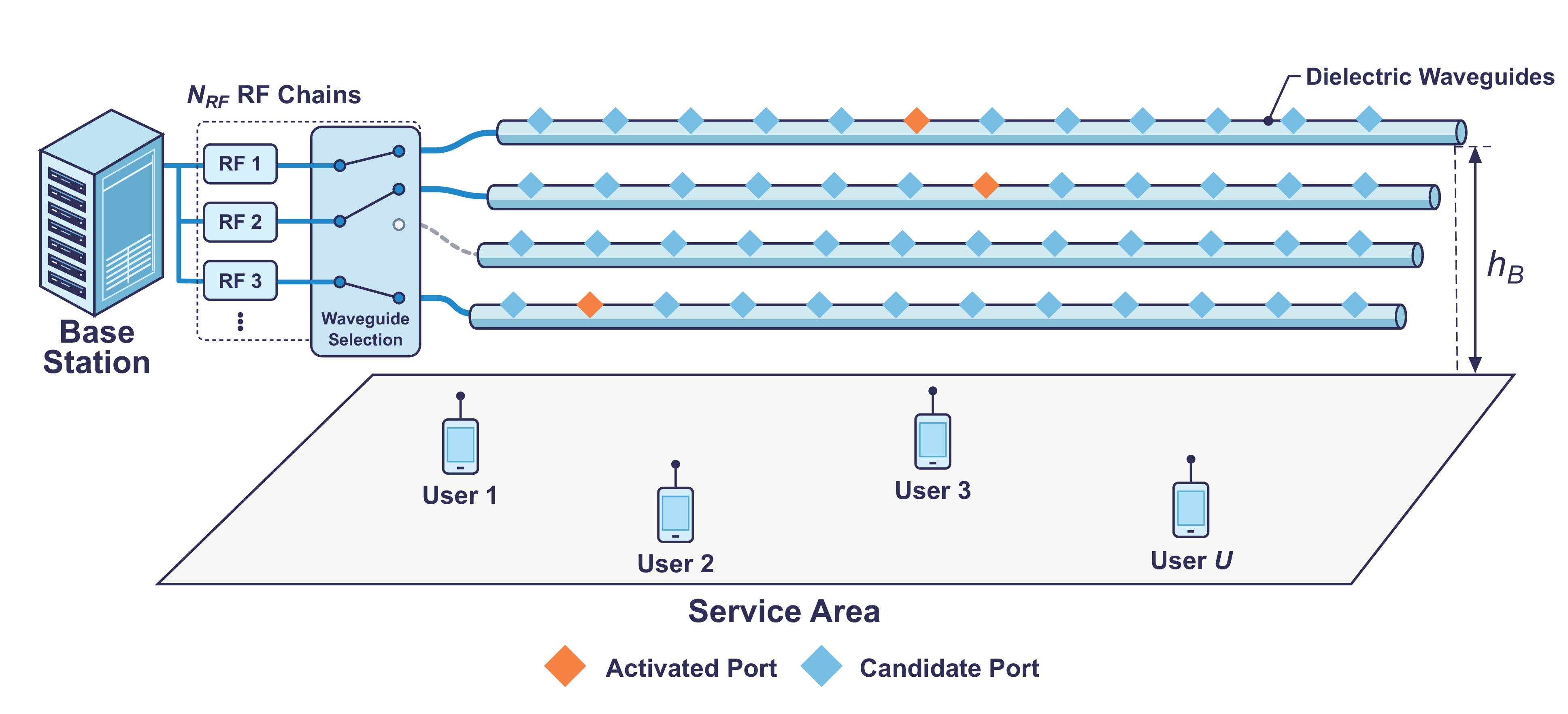}
    \vspace{-0.6cm}
    \caption{Illustration of the  downlink pinching-antenna system.}
    \vspace{-0.2cm}
    \label{fig:system_model}
\end{figure}

As illustrated in Fig.~\ref{fig:system_model}, consider a downlink PASS in which a base station deploys $M$ dielectric waveguides at height $h_{\mathrm B}$ above the service area to serve $U$ single-antenna users.
Each waveguide provides $K$ preconfigured candidate ports, yielding $N=MK$ candidate ports in total.
The BS is equipped with $\Nrf\leq M$ RF chains.
Accordingly, the port selection jointly determines which $\Nrf$ waveguides are selected and which candidate port is selected on each selected waveguide.

Let $\bS=[s_{m,k}]\in\{0,1\}^{M\times K}$ denote the port selection matrix, where $s_{m,k}=1$ indicates that the $k$th port on waveguide $m$ is activated. It is assumed that at most one port can be activated on each waveguide, and  $\Nrf\leq M$ ports are selected in total, where \(\Nrf\) denotes the number of RF chains \cite{xu2026environmentawarenetworkleveldesignpinchingantenna}.
To put it into context, the port selection matrix $\bS$ belongs to the following set
\begin{equation*}
\mathcal{B}\!:= \!
\Big\{
\bS\in \{ 0,1 \}^{M\times K} |
\sum_{k=1}^{K}s_{m,k}\leq 1,\ \forall m,\ 
\sum_{m=1}^{M}\sum_{k=1}^{K}s_{m,k} \! = \!\Nrf \Big\}.
\label{eq:selection_constraints}
\end{equation*}

Following the spherical-wave LoS channel model in \cite{ding2024flexible,wang2025modeling}, let $\bH\in\mathbb C^{U\times MK}$ denote the channel matrix from all candidate ports to all users.
Its $u$th row corresponds to the channel of user $u$, and its $((m-1)K+k)$th column corresponds to the $k$th candidate port on waveguide $m$. The corresponding channel coefficient is
\begin{equation*}
    [\bH]_{u,(m-1)K+k} \! = \!\frac{\sqrt{\beta_0}}{d_{u,m,k}}
    \! \exp\!\left(-j\frac{2\pi}{\lambda_c}d_{u,m,k}\right) 
   \! \exp\!\left(-j\frac{2\pi}{\lambda_g}\ell_{m,k}\right),
\label{eq:channel}
\end{equation*}
where $\beta_0$ denotes the channel power gain at the reference distance, $d_{u,m,k}$ is the distance from candidate port $(m,k)$ to user $u$, and $\ell_{m,k}$ is the propagation distance from the feed point to that port along waveguide $m$, with the feed point referring to the point at which the signal enters the waveguide.
Here, $\lambda_c$ is the carrier wavelength, and $\lambda_g=\lambda_c/n_{\mathrm{eff}}$ is the guided wavelength, with $n_{\mathrm{eff}}$ denoting the effective refractive index of the waveguide.

Given a selection $\bS\in\mathcal B$, denote the activated ports by $(m_r,k_r)$, $r=1,2, \ldots,\Nrf$.
The effective channel matrix $\bH(\bS)\in\mathbb C^{U\times\Nrf}$ is formed by  the corresponding columns of $\bH$, i.e.,
\begin{equation*}
    [\bH(\bS)]_{u,r}
    =
    [\bH]_{u,(m_r-1)K+k_r}
    \label{eq:selected_channel}
\end{equation*}
for $     u=1,2,\ldots,U,\  r=1,2,\ldots,\Nrf.$
Let $\bh_u^H(\bS)$ denote the $u$th row of $\bH(\bS)$. The information symbol $x_u$ intended for user $u$ is precoded by $\bw_u\in\mathbb{C}^{\Nrf}$, with $\mathbb{E}[|x_u|^2]=1$. The received signal at user $u$ is therefore
\begin{equation}\label{eq:received_signal}
  y_u = \bh_u^{H}(\bS)\bw_u x_u + \sum_{v\neq u} \bh_u^{H}(\bS)\bw_v x_v + n_u,
\end{equation}
where $n_u\sim\mathcal{CN}(0,\sigma^2)$ denotes the additive noise.
The multiuser sum rate under the PASS system can be expressed as
\begin{equation}
  R(\bS,\bW) = \sum_{u=1}^{U} \log_2\!\left( 1+ \frac{|\bh_u^{H}(\bS)\bw_u|^2 }{ \sum_{v\neq u} |\bh_u^{H}(\bS)\bw_v|^2+\sigma^2}\right),
  \label{eq:sumrate}
\end{equation}
where $\bW=[\bw_1,\bw_2,\ldots,\bw_U]\in\mathbb{C}^{\Nrf\times U}$.

Our design goal is to jointly optimize the port selection and precoder design, aiming for enhanced sum rate performance, which is formulated as
\begin{equation}
  \begin{aligned}
    \max_{\bS,\bW}
   R(\bS,\bW)  \quad \mathrm{s.t. } 
    ~ \bS\in {\cal B} ,~ \|\bW\|_{F}^{2}\leq P,
  \end{aligned}
  \label{eq:problem}
\end{equation}
where $P$ denotes the total transmit-power budget.
Problem \eqref{eq:problem} is a nonconvex MINLP that couples port selection with continuous precoder design. In particular, the number of feasible port selection realizations is
$
|\mathcal{B}|=\binom{M}{\Nrf}K^{\Nrf},
$
which may grow exponentially with the number of RF chains due to the factor $K^{\Nrf}$, making exhaustive search computationally
prohibitive.

\section{Discrete Diffusion-Based Optimization}
\label{sec:LFDD}

\subsection{Diffusion-Based Optimization Formulation}
\label{ssec:overview}

\begin{figure*}[!t]
    \centering
    \includegraphics[width=\textwidth]{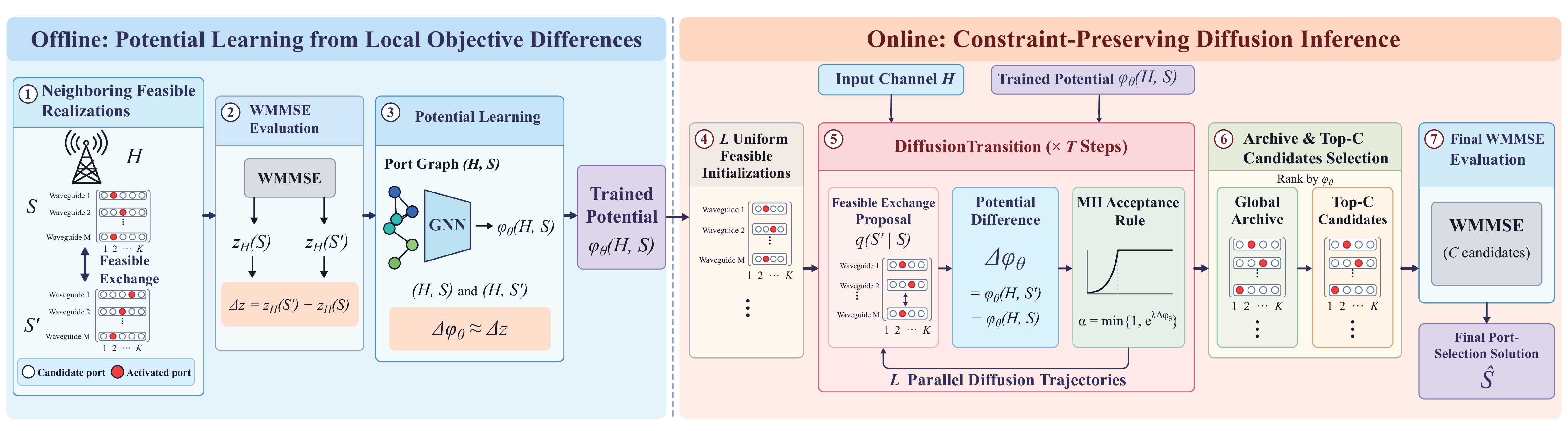}
    \vspace{-0.6cm}
    \caption{Overview of the proposed discrete diffusion framework.}
    \vspace{-0.2cm}
    \label{fig:LFDD}
\end{figure*}

We propose to tackle the design optimization problem by repurposing diffusion models---cutting-edge generative models in machine learning \cite{NEURIPS2020_4c5bcfec}, as an optimization method.
The key idea is to transform the optimization problem \eqref{eq:problem} into a sampling problem from a target distribution, where the sampling process is guided by our dedicated problem-specific potential function. 
Our intention is to leverage the superior sampling capacity to generate high-quality solutions to the design problem \eqref{eq:problem}.
To this end, we rewrite the problem \eqref{eq:problem} as 
\begin{equation}
  \begin{aligned}
  \max_{\bS} & ~f_{\bH}(\bS) := \max_{\|\bW\|_{F}^{2}\leq P} R(\bS,\bW) \\
    \mathrm{s.t.}  & ~ \bS\in {\cal B} ,
  \end{aligned}
  \label{eq:problem2}
\end{equation}
and construct  the corresponding sampling task as 
\begin{equation}
\mbox{sample~} \bS \sim \pi_\lambda(\bS|\bH)  :=  \frac{ \exp\!\left(\lambda f_{\bH}(\bS)\right) }{  \sum_{\widetilde{\bS}\in\setB} \exp\!\left(\lambda f_{\bH}(\widetilde{\bS})\right) },  \label{eq:learned_distribution}
\end{equation}
where $\lambda>0$ is a parameter; when $\lambda\rightarrow +\infty$, the distribution $\pi_\lambda(\bS|\bH)$ becomes a spike around the optimal solution to problem \eqref{eq:problem2} and a near-optimal solution can be sampled with high probability. Therefore, the combinatorial optimization problem can be approached by sampling high-probability realizations from $\pi_\lambda(\bS|\bH)$, motivating a discrete diffusion-based sampling procedure.
 
However, directly applying such a diffusion-based sampling procedure to the considered problem faces several challenges: First, the function $f_{\bH}(\bS)$ in the target distribution $\pi_\lambda(\bS|\bH)$ requires solving the nonconvex sum rate optimization problem for each realization of $\bS$, rendering high computational complexity and sensitivity to different initializations.
Second, in order to obtain an accurate approximation of the target distribution by direct supervised learning approaches, a large number of optimal or near-optimal port selection solutions are required as training labels. 
Third and more importantly, usually the diffusion process is unconstrained, but our variable $\bS$ is subject to a stringent combinatorial constraint.

Therefore, we exploit the problem structure and propose a lightweight discrete diffusion method that does not rely on optimal-solution labels and preserves the combinatorial structure of $\setB$ throughout the diffusion process.
An overview of the proposed framework is illustrated in Fig.~\ref{fig:LFDD}.
Our development is as follows.

\subsection{Potential Learning from Local Objective Differences}
\label{ssec:potential_learning}

Given any channel instance $\bH$, evaluating the function $f_{\bH}(\bS)$ in \eqref{eq:problem2} and also the distribution $\pi_\lambda(\bS|\bH)$ requires solving the associated inner nonconvex precoder-design problem in  $f_{\bH}(\bS)$ from scratch, usually by the well-established WMMSE algorithm \cite{shi2011iteratively} with a per-iteration complexity $\mathcal{O}(U\Nrf^2+\Nrf^3)$ for a high-quality solution. If we evaluate each selection using the sum rate achieved by a prescribed WMMSE procedure $\hat{f}_{\bH}(\bS)$, this will incur considerable computational cost in generating optimal-solution labels and evaluating candidate realizations during sampling.
To resolve this issue, we propose to use a surrogate distribution $ \tilde{\pi}_\theta(\bS|\bH)$, driven by a graph neural network $\phi_\theta(\bH,\bS)$ parameterized by $\theta$, i.e.,
\begin{equation}  \label{eq:surrogate_distribution}
  \tilde{\pi}_\theta(\bS|\bH)
  \propto \exp( \lambda\phi_\theta(\bH,\bS)),
\end{equation}  
to globally approximate $\pi_\lambda(\bS|\bH)$.
We expect that the learned $\tilde{\pi}_\theta(\bS|\bH)$ has the generalization capacity to predict $\pi_\lambda(\bS|\bH)$ for unseen $\bH$  to avoid intensive WMMSE evaluations during sampling.

The coming question is to learn the potential function $\phi_\theta(\bH,\bS)$ to approximate $\hat{f}_{\bH}(\bS)$.
Instead of learning the absolute $\hat{f}_{\bH}(\bS)$, we turn to learn its relative landscape via pairwise realizations to better harness the relative relations among various port selection realizations. 
Consider two neighboring realizations $\bS,\bS'\in\setB$. Taking the ratio of their probabilities yields
\begin{equation}
  \log
  \frac{\pi_\lambda(\bS'|\bH)}
  {\pi_\lambda(\bS|\bH)}= \lambda[
  \hat{f}_{\bH}(\bS')-\hat{f}_{\bH}(\bS)].
  \label{eq:target_local_ratio}
  \end{equation}
The above relative difference is also beneficial to reduce the impact of different channel scales. To further adapt to a broad range of channel distributions, we normalize the objective as
\begin{equation}
  z_{\bH}(\bS)
  =
  \frac{\hat{f}_{\bH}(\bS)}{B(\bH)},
  \quad
  B(\bH)
  =
  \sum_{u=1}^{U}
  \log_2\!\left(
  1+\frac{P}{\sigma^2}\|\bh_u\|_2^2
  \right),
  \label{eq:normalized_value}
\end{equation}
where $B(\bH)>0$ is independent of $\bS$, and is an upper bound of the sum rate $R(\bS, \bW)$.
Thus, we define the normalized relative difference as
\begin{equation}
  \begin{aligned}
  \Delta z
  &\triangleq
  z_{\bH}(\bS')-z_{\bH}(\bS).
  \end{aligned}
  \label{eq:potential_difference}
\end{equation}

Correspondingly, we construct from  the surrogate distribution 
  \begin{equation}
  \log
  \frac{\tilde{\pi}_\theta(\bS'|\bH)}
  {\tilde{\pi}_\theta(\bS|\bH)}
  =  \lambda[
  \phi_\theta(\bH,\bS')
  - \phi_\theta(\bH,\bS)].
  \label{eq:surrogate_local_ratio}
\end{equation}
and define
\begin{equation}\label{eq:norm_poten}
    \Delta\phi_\theta
 \triangleq
  \phi_\theta(\bH,\bS')-\phi_\theta(\bH,\bS).
\end{equation}
The training objective is to minimize the difference between  \eqref{eq:potential_difference} and  \eqref{eq:norm_poten} via minimizing the following Huber loss
\begin{equation}
  \mathcal{L}_{\mathrm{edge}}
  =
  \operatorname{Huber}_{\delta}
  \!\left(
  \Delta\phi_\theta-\Delta z
  \right),
  \label{eq:loss_edge}
\end{equation}
where $\delta$ is the Huber threshold \cite{Huber1992}.
The training process requires $\hat{f}_{\bH}(\bS)$ obtained from   WMMSE as an oracle for a set of channel instances. 
This should be distinguished from solution-supervised learning for problem \eqref{eq:problem2}, which requires such solution labels.

As shown in Steps~1--3 of Fig.~\ref{fig:LFDD}, neighboring feasible realizations are constructed (Step~1), evaluated by WMMSE (Step~2), and used for potential learning (Step~3).

\subsection{Constraint-Preserving Diffusion Inference}
\label{ssec:diffusion}

Next, we address the challenge of preserving the combinatorial constraints throughout the diffusion process. 
In conventional continuous diffusion models, the diffusion process between the target distribution and the standard Gaussian distribution is realized by gradually adding/removing Gaussian noise and lies in a continuous space. 
In our discrete diffusion, this diffusion process is implemented through Markov transitions over a discrete state space between the target distribution and the uniform distribution \cite{austin2021structured}. 
If there were no selection constraint $\cal B$, the discrete diffusion model has been studied in \cite{sun2023difusco} in a ``free'' space.
But the method there cannot be directly extended to handle the selection constraint $\cal B$ in our problem.

Motivated by the constrained discrete diffusion idea \cite{cardei2025constraineddiscretediffusion}, we construct a constraint-preserving forward diffusion process directly over the feasible set $\setB$. 
Starting from a feasible realization $\bS\in\setB$, we uniformly select and deactivate one active port, then uniformly activate a legal port other than the one just removed, yielding a neighboring realization $\bS'\in\setB$. 
Since the transition is defined entirely through feasible exchanges, every intermediate realization remains in $\setB$ by construction.
Let 
\[
q(\bS'|\bS)=1/\{\Nrf[K(M-\Nrf+1)-1]\}
\]
denote the  probability of transition from $\bS$ to $\bS'$. 
The transition is symmetric, i.e., $q(\bS'|\bS)=q(\bS|\bS')$.
Thus, the above random transition maintains $\bS$ in $\setB$, and progressively transforms the distribution to a uniform stationary distribution $\pi_0(\bS)=1/|\setB|$ for $\bS\in\setB$.

During the sampling process for generating candidate realizations, we aim to assign higher probabilities to those with larger sum rates.
However, direct sampling from the learned surrogate distribution is generally intractable.
We therefore employ a diffusion process to generate candidate realizations and an MH rule to determine their acceptance.
Higher-objective candidates are favored, while lower-objective ones retain a nonzero acceptance probability to preserve exploration.
The MH acceptance probability depends only on the relative probability ratio between two solutions, as defined in \eqref{eq:surrogate_local_ratio}.
This property naturally motivates our strategy of learning relative objective differences rather than absolute objective values in Section~\ref{ssec:potential_learning}.
Accordingly, the acceptance ratio for a candidate realization $\bS$ is defined as
\begin{equation}
  p_{\sf acc}(\bS_{\sf curr},\bS)
  =
  \min\!\left\{
  1,
  \exp\!\left(
  \lambda\Delta\phi_\theta
  \right)
  \right\},
  \label{eq:mh}
\end{equation}
where $\bS_{\sf curr}$ is the current selection stage.

To improve the sampling efficiency, we run $L$ diffusion trajectories in parallel, each initialized from the uniform distribution over $\setB$. 
We select top-$C$ candidates ranked by the learned potential $\phi_\theta(\bH,\bS)$, and finally select the one with the highest sum rate as the final port selection solution to problem \eqref{eq:problem2}.
During the diffusion inference stage, only $C$ candidates require WMMSE evaluation. 
In contrast, greedy search requires $K\sum_{r=0}^{\Nrf-1}(M-r)$ WMMSE evaluations, while Beam-$B$ requires $KM+BK\sum_{r=1}^{\Nrf-1}(M-r)$ under the same sequential expansion strategy \cite{cheng2026fastpinchingantennaactivationaircomp}. 

As shown in Steps~4--7 of Fig.~\ref{fig:LFDD}, we initialize $L$ feasible realizations (Step~4), perform constraint-preserving diffusion transitions (Step~5), select the top-$C$ candidates from the archive (Step~6), and finally evaluate them by WMMSE (Step~7).

\section{Simulation Results}
\label{sec:exp}

\subsection{Simulation Setup}
\label{ssec:setup}

We consider a $28\,\mathrm{GHz}$ PASS over a $30\,\mathrm{m}\times30\,\mathrm{m}$ service area, with $n_{\mathrm{eff}}=1.44$, waveguides at height $3\,\mathrm{m}$, and noise power $\sigma^2=10^{-12}\,\mathrm{W}$.
The channels are generated randomly based on \eqref{eq:channel}.  
Users are placed around a uniformly sampled cluster center via isotropic Gaussian displacements with standard deviation $2.4\,\mathrm{m}$, excluding locations outside the area or within $0.5\,\mathrm{m}$ of any port.
The small- and large-scale settings are $(M,K,\Nrf,U)=(6,4,4,4)$ and $(24,8,16,16)$, with $(L,T,C)=(64,50,8)$ and $(128,400,64)$, respectively. The $L$ diffusion trajectories run in parallel on an NVIDIA RTX 4090 D GPU.
We compare the proposed method with greedy search and beam search of width two (Beam-2), using exhaustive search as the small-scale reference.
Greedy search retains the best feasible partial selection at each step, whereas Beam-2 retains the best two.
All methods use the same WMMSE implementation, a $2000$-iteration limit, and relative tolerance $10^{-6}$.

The potential $\phi_\theta(\bH,\bS)$ is implemented by a GNN with three
message-passing layers, hidden dimension $128$, SiLU activations,
and layer normalization, totaling $551{,}809$ trainable parameters.
Port embeddings are aggregated at the waveguide and global levels
and combined with user-correlation features to predict the potential.
Starting from small-scale pretrained weights, the network is further
trained on $1024$ large-scale channels for $26$ epochs using Adam
with learning rate $10^{-4}$ and batch size $8$.
Each channel visit samples $16$ neighboring pairs.
The Huber loss has a transition point of $0.02$, while auxiliary
value-regression and ranking losses each have weight $0.1$.
The same weights are used for both system sizes without test-time updates.
Results are averaged over $100$ independent test channels.
Gap and gain are computed per channel relative to exhaustive and greedy search, respectively, and then averaged.


\subsection{Performance Comparison}
\label{ssec:main}

Table~\ref{tab:all_algorithms_power2025} shows that, at $20$ and $25\,\mathrm{dBm}$, the proposed method achieves gaps of only $0.30\%$ and $0.24\%$ to the exhaustive reference, respectively. Meanwhile, its runtime is reduced from $60.240$ s and $86.725$ s for exhaustive search to only $0.394$ s and $0.363$ s, corresponding to more than $150\times$ speedup.
Compared with greedy search, the proposed method is approximately $6.9\times$ and $9.2\times$ faster while also achieving higher sum rates. It also outperforms Beam-2 in both sum rate and runtime, showing that the proposed diffusion method can approach exhaustive-search performance without requiring extensive candidate evaluation.

\begin{table}[!t]
\centering
\caption{Average performance over 100 test channels;
$(M,K,\Nrf,U)=(6,4,4,4)$.}
\label{tab:all_algorithms_power2025}
	\renewcommand{\arraystretch}{1.}
	\resizebox{\linewidth}{!}{%
\begin{tabular}{M{14mm}|M{10mm}M{12mm}M{12mm}|M{10mm}M{12mm}M{12mm}}
\toprule
\multirow{2}{*}{\textbf{Method}}
& \multicolumn{3}{c|}{\textbf{Power = 20 dBm}}
& \multicolumn{3}{c}{\textbf{Power = 25 dBm}} \\
\cmidrule(lr){2-4}\cmidrule(lr){5-7}
& Rate & Gap (\%) & Time (s) & Rate & Gap (\%) & Time (s) \\
\midrule
Exhaustive & 35.665 & 0.00 & 60.240 & 42.292 & 0.00 & 86.725 \\
Greedy & 33.882 & 5.00 & 2.708 & 38.648 & 8.58 & 3.335 \\
Beam-2 & 34.685 & 2.77 & 4.014 & 39.784 & 5.87 & 5.549 \\
Proposed & \textbf{35.564} & \textbf{0.30} & \textbf{0.394}
& \textbf{42.194} & \textbf{0.24} & \textbf{0.363} \\
\bottomrule
\end{tabular}
}
\end{table}

In the large-scale setting, Table~\ref{tab:all_algorithms_power2025_large} shows mean sum rate gains of $5.09\%$ and $4.90\%$ over greedy search, with $18.8\times$ and $14.4\times$ speedups at $20$ and $25\,\mathrm{dBm}$, respectively.
Compared with Beam-2, the proposed method achieves $2.16\%$ and $2.34\%$ higher sum rates with $37.0\times$ and $26.9\times$ speedups, respectively.
This efficiency mainly stems from using only $64$ terminal WMMSE evaluations, whereas greedy search requires $2112$ candidate evaluations during expansion.

\begin{figure}[!t]
    \centering
    \begin{subfigure}[b]{0.48\columnwidth}
        \centering
        \includegraphics[width=\textwidth]{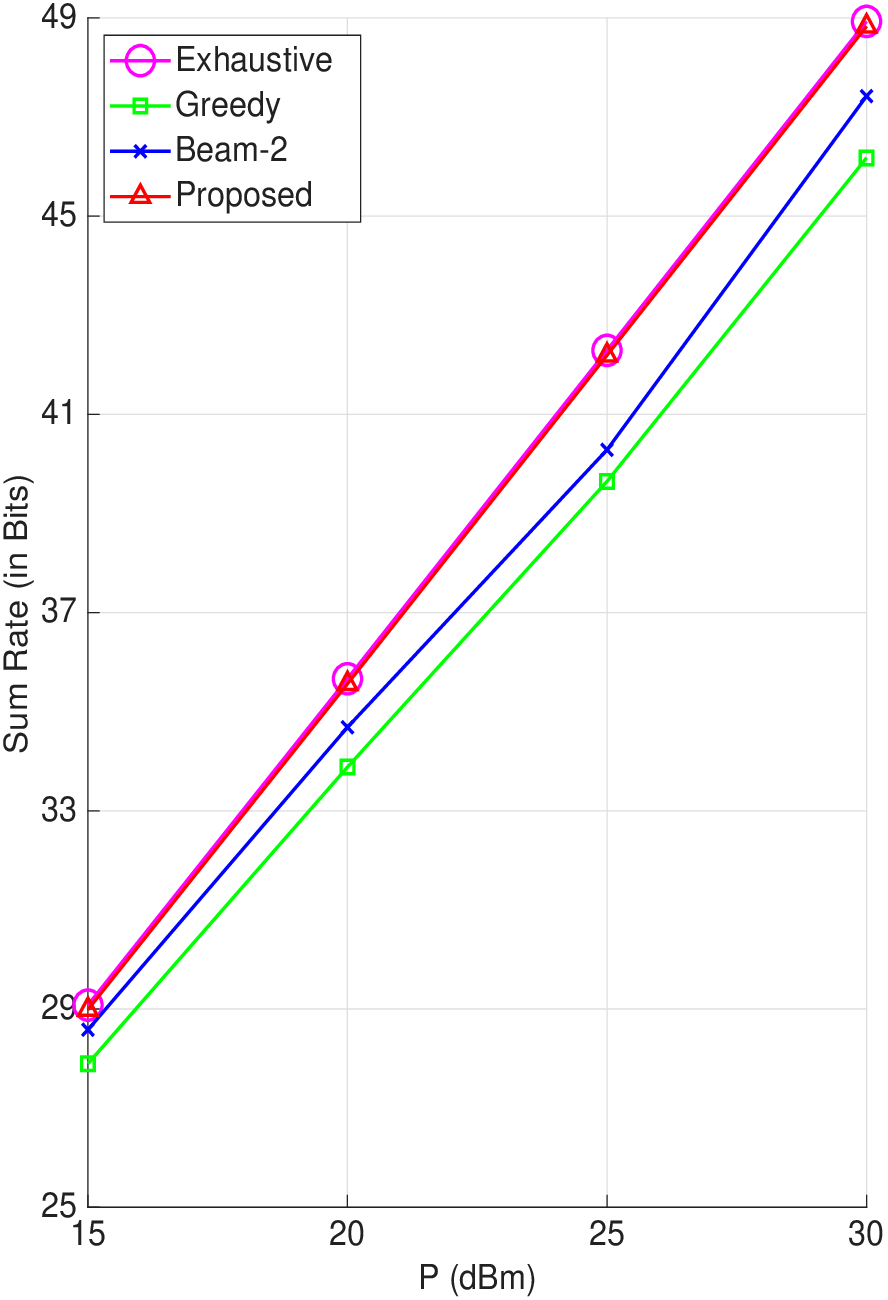}
        \caption{\tiny $(M,K,\Nrf,U)=(6,4,4,4)$.}
        \label{fig:small}
    \end{subfigure}
    \hfill
    \begin{subfigure}[b]{0.48\columnwidth}
        \centering
        \includegraphics[width=\textwidth]{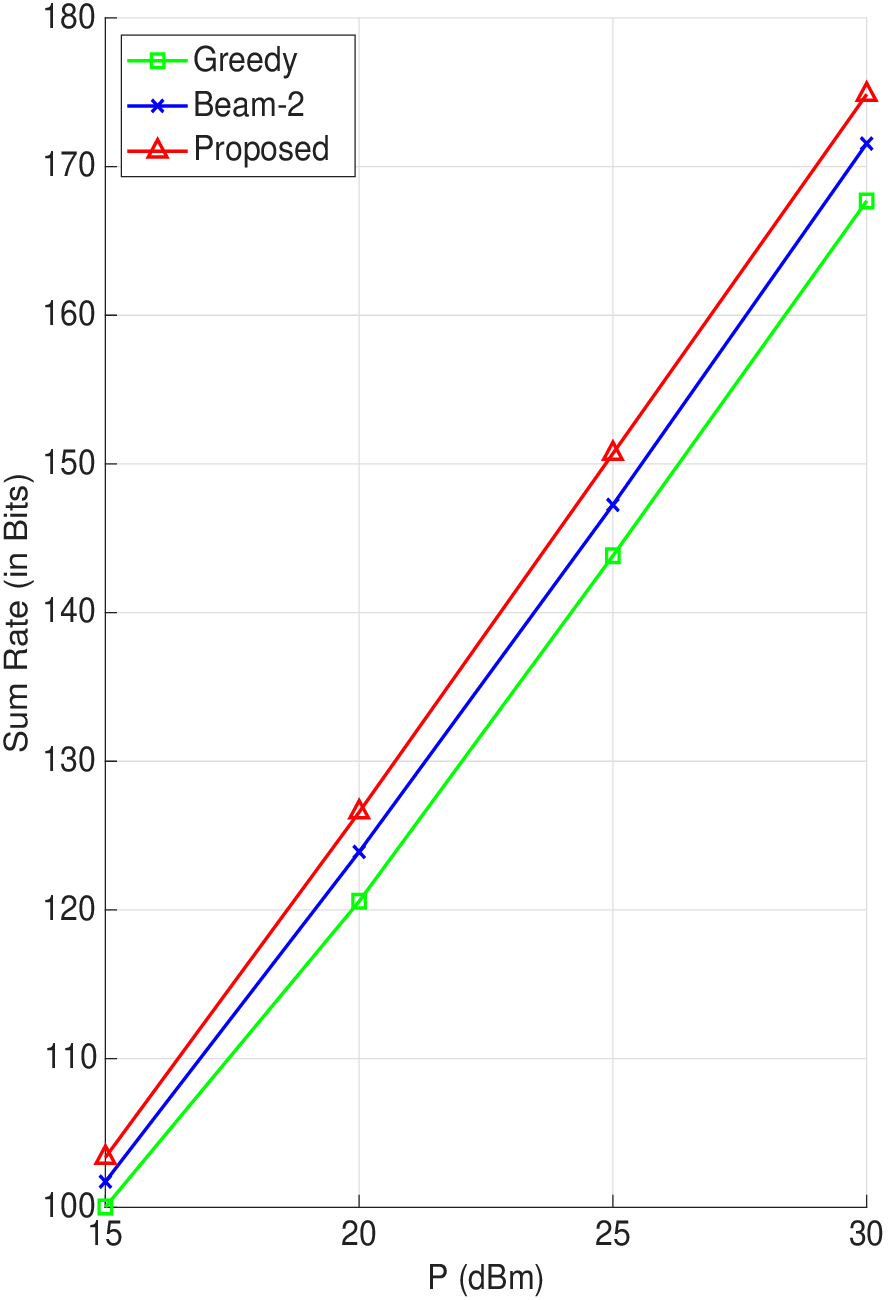}
        \caption{\tiny $(M,K,\Nrf,U)=(24,8,16,16)$.}
        \label{fig:large}
    \end{subfigure}
    
    \caption{Average sum rate versus transmit power.}
    \vspace{-1.5cm}
    \label{fig:sumrate_power}
\end{figure}

Fig.~\ref{fig:sumrate_power} shows consistent trends across the
considered transmit-power range: the proposed method closely
tracks exhaustive search in the small-scale setting and achieves
higher sum rates than greedy search and Beam-2 in the large-scale
setting.

\begin{table}[!t]
\centering
\caption{Average performance over 100 test channels;
$(M,K,\Nrf,U)=(24,8,16,16)$.}
\label{tab:all_algorithms_power2025_large}
	\renewcommand{\arraystretch}{1.1}
	\resizebox{\linewidth}{!}{%
\begin{tabular}{M{11mm}|M{10mm}M{12mm}M{11mm}|M{10mm}M{12mm}M{11mm}}
\toprule
\multirow{2}{*}{\textbf{Method}}
& \multicolumn{3}{c|}{\textbf{Power = 20 dBm}}
& \multicolumn{3}{c}{\textbf{Power = 25 dBm}} \\
\cmidrule(lr){2-4}\cmidrule(lr){5-7}
& Rate & Gain (\%) & Time (s) & Rate & Gain (\%) & Time (s) \\
\midrule
Greedy & 120.580 & -- & 142.800 & 143.823 & -- & 195.826 \\
Beam-2 & 123.906 & 2.85 & 280.724 & 147.248 & 2.49 & 366.060 \\
Proposed & \textbf{126.580} & \textbf{5.09} & \textbf{7.590}
& \textbf{150.698} & \textbf{4.90} & \textbf{13.620} \\
\bottomrule
\end{tabular}
}
\end{table}

\section{Conclusion}
\label{sec:conclusion}

We studied joint port selection and precoder design for sum rate maximization in multiuser PASS systems.
Our key idea is to recast the MINLP problem as sampling from a target distribution over port selection solutions.
We learned a potential function from relative objective differences, without requiring optimal port selection solutions as training labels.
During sampling, the diffusion process preserves the selection constraints, while a guided MH rule steers the search toward high-quality solutions.
Simulations demonstrated competitive sum rate performance with low inference complexity.

\pagebreak
 
\bibliographystyle{IEEEbib}
\bibliography{strings,refs}

\end{document}